\documentclass{aa}  

\usepackage{graphicx}
\usepackage{txfonts}
\usepackage{natbib}
\begin{document} 

\title{Constraining the efficiency of angular momentum transport with asteroseismology of red giants: the effect of stellar mass}
\titlerunning{Internal angular momentum transport in red giants}

\author{P. Eggenberger\inst{1}
\and N. Lagarde\inst{2}
\and A. Miglio\inst{3} 
\and J. Montalb\'{a}n\inst{4} 
\and S. Ekstr\"om\inst{1}
\and C. Georgy\inst{1}
\and G. Meynet\inst{1}
\and S. Salmon\inst{5}
\and T.~Ceillier\inst{6}
\and R. A.~Garc\'{i}a\inst{6}
\and S. Mathis\inst{6}
\and S. Deheuvels\inst{7}
\and A. Maeder\inst{1}
\and J. W. den Hartogh\inst{8}
\and R. Hirschi\inst{8,9}}

\institute{Observatoire de Gen\`eve, Universit\'e de Gen\`eve, 51 Ch. des Maillettes, CH-1290 Sauverny, Suisse \\	
\email{patrick.eggenberger@unige.ch}
\and
Institut Utinam, CNRS UMR 6213, Universit\'e de Franche-Comt\'e, OSU THETA Franche-Comt\'e-Bourgogne, Observatoire de Besan\c{c}on, 25010, Besan\c{c}on, France
\and
School of Physics and Astronomy, University of Birmingham, Edgbaston, Birmingham B15 2TT, UK
\and
Dipartimento di Fisica e Astronomia, Universit\`a di Padova, Vicolo dell'Osservatorio 3, I-35122 Padova, Italy
\and
D\'epartement d'Astrophysique, G\'eophysique et Oc\'eanographie, Universit\'e de Li\`ege, All\'ee du 6 Ao\^ut 17, 4000 Li\`ege, Belgium
\and
Laboratoire AIM, CEA/DRF-CNRS, Universit\'e Paris 7 Diderot, IRFU/SAp, Centre de Saclay, 91191, Gif-sur-Yvette, France
\and
Universit\'e de Toulouse, UPS-OMP, IRAP, 31400, Toulouse, France
\and
Astrophysics group, Lennard-Jones Laboratories, Keele University, ST5 5BG, Staffordshire, UK
\and
Kavli Institute for the Physics and Mathematics of the Universe (WPI), University of Tokyo, 5-1-5 Kashiwanoha, Kashiwa, 277-8583, Japan}

\date{Received; accepted}

% \abstract{}{}{}{}{} 
% 5 {} token are mandatory

 \abstract
   {Constraints on the internal rotation of red giants are now available thanks to asteroseismic observations. Preliminary comparisons with rotating stellar models indicate that an undetermined additional process for the internal transport of angular momentum is required in addition to purely hydrodynamic processes.}
   {We investigate how asteroseismic measurements of red giants can help us characterize the additional transport mechanism.}
   {We first determine the efficiency of the missing transport mechanism for the low-mass red giant KIC~7341231 by computing rotating models that include an additional viscosity corresponding to this process. We then discuss the change in the efficiency of this transport of angular momentum with the mass, metallicity and evolutionary stage in the light of the corresponding viscosity determined for the more massive red giant KIC~8366239.}
   {In the case of the low-mass red giant KIC~7341231, we find that the viscosity corresponding to the additional mechanism is constrained to the range $\nu_{\rm add} = 1 \times 10^{3}$ -- $1.3 \times 10^{4} $\,cm$^2$\,s$^{-1}$. This constraint on the efficiency of the unknown additional transport mechanism during the post-main sequence is obtained independently of any specific assumption about the modelling of rotational effects during the pre-main sequence and the main sequence (in particular, the braking of the surface by magnetized winds and the efficiency of the internal transport of angular momentum before the post-main-sequence phase). When we assume that the additional transport mechanism is at work during the whole evolution of the star together with a solar-calibrated braking of the surface by magnetized winds, the range of $\nu_{\rm add}$ is reduced to $1$ --  $4 \times 10^{3}$\,cm$^2$\,s$^{-1}$. In addition to being sensitive to the evolutionary stage of the star, we show that the efficiency of the unknown process for internal transport of angular momentum increases with the stellar mass.}
{}



   \maketitle
%
%________________________________________________________________

\section{Introduction}
\label{intro}

The inclusion of hydrodynamic transport processes that are due to rotation in a stellar evolution code can lead to significant changes in the global and internal properties of stellar models \citep[e.g.][]{mae09}. The changes induced by rotation are found to be sensitive to the prescriptions used for the modelling of these dynamical processes, and in particular to the modelling of the transport of chemical elements and angular momentum by meridional circulation and shear instabilities \citep[e.g.][]{mey13}. We are still far from having a clear understanding of the various transport processes at work in stellar interiors. Direct observations of internal properties of stars, and in particular of internal rotation rates, are thus extremely valuable to progress in the modelling of these processes. Measurements of rotational splittings of solar-like oscillations for stars with different masses and at different evolutionary stages are especially promising.

The rotation profile of the Sun is a first important constraint for stellar models that take the internal transport of angular momentum into account. From measurements of rotational splittings of solar pressure modes, an approximately flat rotation profile is deduced in the solar radiative zone \citep[see e.g.][]{bro89,els95,kos97,cou03,gar07}. Rotating models of the Sun that include the transport of angular momentum by hydrodynamic processes due to rotation predict a rapidly rotating core at the solar age \citep[e.g.][]{pin89,cha95,cha05,egg05_mag,tur10}. This discrepancy with helioseismic data indicates that current prescriptions for meridional currents and shear instability are not able to produce a sufficent coupling in the solar case. An efficient mechanism for angular momentum transport is thus required in the radiative zone of the Sun.

Motivated by the wealth of information provided by the solar five-minute oscillations, solar-like oscillations have now been detected and characterized for a large number of stars. While it is difficult to obtain information about the internal rotation of main-sequence solar-type stars from asteroseismic measurements of pressure modes, which mainly propagate in the external stellar layers \citep[see e.g.][]{lun14,ben15,nie15}, constraints on the core rotation rates can be obtained for red giants. Mixed oscillation modes, which behave similarly to pressure modes in the external stellar layers and similarly to gravity modes in the stellar center, can indeed be observed for these evolved stars. With the launch of the {\it Kepler} spacecraft \citep{bor10}, information about the internal rotation are now available for red giants \cite[][]{bec12, deh12, mos12, deh14, deh15, dim16}.

Rotational frequency splittings of mixed modes have first been measured for the red giant KIC~8366239 \citep{bec12}. Based on these data, \cite{bec12} deduced that solid-body rotation can be ruled out and that the core of this star is rotating at least ten times faster than its surface. Stellar models including rotational effects predict fast-rotating cores during the red giant phase as a result of the rapid central contraction occurring after the main sequence \citep[e.g.][]{pal06,egg10_rg}. However, rotating models of KIC~8366239 predict too steep rotation profiles compared to asteroseismic constraints. Similarly to the solar case, one concludes that an efficient mechanism for the internal transport of angular momentum is at work during the post-main-sequence phase in addition to hydrodynamic processes\footnote{hydrodynamic processes refer here only to mechanisms related to rotation, namely the transport of angular momentum by meridional circulation and shear instabilities.} \citep{egg12_rg, mar13, can14}. 

This conclusion was obtained for the red giant KIC~8366239, which is massive enough (mass of about 1.5\,M$_{\odot}$) to exhibit a convective core during the main sequence. \cite{deh12} deduced the core rotation rate of the low-mass red giant KIC~7341231 (mass of about 0.84\,M$_{\odot}$) and an upper limit for its surface rotation rate. Comparison with models of KIC~7341231 computed with shellular rotation shows that an efficient mechanism for angular momentum transport is also required in addition to hydrodynamic processes for this low-mass red giant, which has a radiative core during the main sequence \citep{cei12,cei13}.

While asteroseismic observations point toward the need for an efficient transport of angular momentum in stellar radiative zones of evolved stars, the physical nature of such a process remains an open question. As for the Sun, magnetic fields could play an important role for the internal transport of angular momentum. The impact of the Tayler-Spruit dynamo \citep{spr99,spr02} on the internal rotation of red giants has been investigated by \cite{can14}. This study indicates that this mechanism does not provide a sufficient coupling to correctly reproduce the low values of core rotation rates deduced for red giants. A preliminary study of the effects of internal gravity waves on the transport of angular momentum in the interior of red giants also suggests that these waves are not able to extract angular momentum from the stellar core \citep{ful14}. The transport of angular momentum by mixed oscillation modes has also been investigated \citep{bel15b}. \cite{bel15b} showed that this transport seems to play a negligible role during the subgiant and the early red-giant phase, but could be important for more evolved red giants. 

To reveal the physical nature of this unknown transport mechanism, one has to characterize its efficiency. This has been done for the 1.5\,M$_{\odot}$ red giant KIC~8366239 by introducing an additional viscosity $\nu_{\rm add}$ corresponding to the unknown process in the equation describing the internal transport of angular momentum. In the case of KIC~8366239, one finds that the mean value for the efficiency of this transport mechanism is strongly constrained by asteroseismic measurements to the value of $\nu_{\rm add}=3 \times 10^{4}$\,cm$^2$\,s$^{-1}$ \citep{egg12_rg}. It is also important to study how the efficiency of this process changes with the global stellar parameters. Comparison with measurements of core rotation rates for a large number of red giants by \cite{mos12} indicates that the value of $\nu_{\rm add}$ must increase when evolution proceeds on the red-giant branch \cite[][]{can14,egg15,spa16}. \cite{spa16} have shown that a value of $\nu_{\rm add}$ that increases during the evolution with the radial rotational shear leads to a good agreement with the core rotation rates of \cite{mos12}. This interesting result has been obtained for a fixed mass of 1.25\,M$_{\odot}$, a fixed value for the initial rotation of the star and by assuming solid-body rotation before the post-main-sequence phase. It is thus interesting to study the impact of the stellar mass on the efficiency of the internal transport of angular momentum for red giants at a similar evolutionary stage. It is also important to investigate how the constraints obtained on $\nu_{\rm add}$ from asteroseismic measurements of red giants are sensitive to the numerous uncertainties related to the modelling of rotational effects before the post-main-sequence phase.

In the present paper, we first investigate how the additional viscosity can be determined for the low-mass red giant KIC~7341231 from the value of the core rotation rate and the upper limit on its surface rotation rate deduced by \cite{deh12}. In particular, we study the impact of the uncertainties related to the modelling of the magnetic braking of the stellar surface and the internal transport of angular momentum during the main sequence on the determination of the efficiency required for this process during the post-main sequence. We then discuss how the mean viscosity associated with the additional transport mechanism changes with the stellar mass and the evolutionary stage.

\section{Additional viscosity needed for KIC~7341231}

\subsection{Physics of the models}

Rotating stellar models are computed with the Geneva stellar evolution code \citep{egg08}. This code includes a comprehensive treatment of shellular rotation \citep{zah92,mae98}. The details of the inclusion of rotational effects in the Geneva code are not repeated here but can be found in Sect.~2 of \cite{egg10_sl}. 

In the continuity of the work done for the red giant KIC~8366239, we introduce a constant viscosity $\nu_{\rm add}$ corresponding to the unknown angular momentum transport process. This viscosity is included in addition to the diffusive transport of angular momentum by the shear instability and the advective transport by meridional currents. Of course, we do not pretend that the physical nature of the unkown additional transport process is such that it should effectively result in a constant viscosity in stellar radiative zones. Introducing this additional viscosity is however a valuable mean to first investigate whether asteroseismic data of red giants are able to provide us with quantitative constraints on the efficiency of this undetermined mechanism. Then one can try to deduce its mean efficiency as a function of global stellar parameters such as the mass and the evolutionary stage. The additional process is only taken into account for the transport of angular momentum, but not for the transport of chemical elements. This is motivated by the solar case, where the approximately uniform rotation in the solar radiative zone indicates a very efficient transport of angular momentum, while the surface abundances of light elements require a low efficiency of rotational mixing. The following equation is then solved to follow the transport of angular momentum in radiative zones:

\begin{equation}
  \rho \frac{{\rm d}}{{\rm d}t} \left( r^{2}\Omega \right)_{M_r} 
  =  \frac{1}{5r^{2}}\frac{\partial }{\partial r} \left(\rho r^{4}\Omega
  U(r)\right)
  + \frac{1}{r^{2}}\frac{\partial }{\partial r}\left(\rho D r^{4}
  \frac{\partial \Omega}{\partial r} \right) \, , 
\label{transmom}
\end{equation}

with $r$ and $\rho$ the characteristic radius and the mean density on an isobar, respectively. $\Omega(r)$ is the mean angular velocity and $U(r)$ denotes the velocity of the meridional circulation in the radial direction. The total diffusion coefficient $D$ takes into account the transport of angular momentum by both the shear instability and the additional viscosity associated with the undetermined transport process: $D=D_{\rm shear} + \nu_{\rm add}$. A constant value of the angular velocity as a function of the radius is assumed in convective zones, in accordance with the solar case.

\begin{figure}[htb!]
\resizebox{\hsize}{!}{\includegraphics{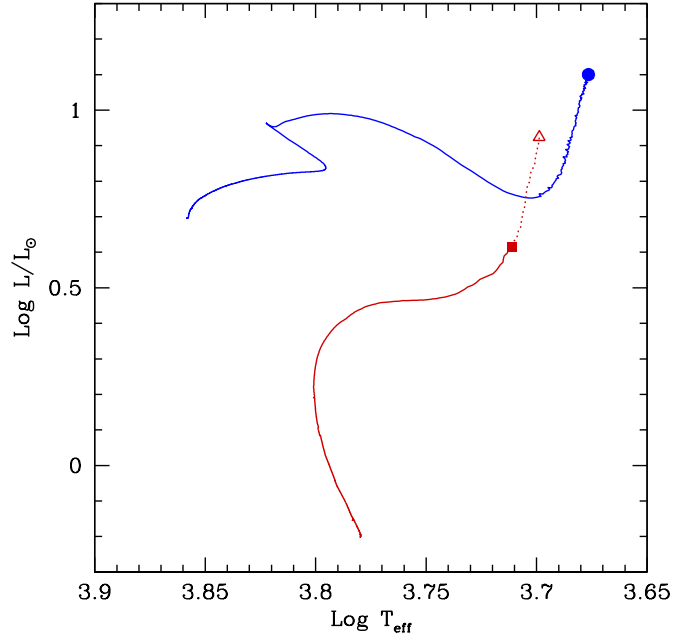}}
 \caption{Evolutionary tracks in the HR diagram for representative rotating models of KIC~7341231 (red lines) and KIC~8366239 (blue line). The square and dot indicate the location of models of KIC~7341231 and KIC~8366239, respectively. The open triangle corresponds to a more evolved model of KIC~7341231, which shares the same value of $\log g$ as the more massive star KIC~8366239.}
  \label{dhr_fig}
\end{figure}

The modelling of KIC~7341231 including the additional viscosity is performed in the continuity of the study by \cite{cei13} that focussed on models computed with rotational effects only. The metallicity is then fixed to $[$Fe/H$] = -1$ with an initial helium mass fraction $Y_{\rm ini} = 0.260$. The solar chemical mixture of \cite{gre93} is used together with a solar-calibrated value for the mixing-length parameter. As in \cite{cei13}, models that simultaneously reproduce the observed values of the large frequency separation and the asymptotic period spacing are considered as representative models of KIC~7341231. In this way, \cite{cei13} determined a mass of 0.84\,M$_{\odot}$ for KIC~7341231 in good agreement with the detailed modelling of KIC~7341231 done by \cite{deh12} using non-rotating models. This also illustrates that the impact of rotational mixing on global parameters is negligible for models of red giants computed with low initial rotation velocities on the zero-age main sequence (ZAMS). Figure~\ref{dhr_fig} shows the location in the HR diagram of a representative model of KIC~7341231 (red square), which corresponds to a 0.84\,M$_{\odot}$ star with a low initial velocity on the ZAMS of 2\,km\,s$^{-1}$. The location of the model corresponding to the more massive target KIC~8366239 is also shown (blue dot in Fig.~\ref{dhr_fig}).

\subsection{Models with an additional viscosity of $3 \times 10^{4}$\,cm$^2$\,s$^{-1}$}

As mentioned in the introduction, a first constraint on the efficiency of the missing angular momentum transport process has been obtained from rotational splittings of mixed oscillation modes for the red giant KIC~8366239 \citep{egg12_rg}. A viscosity of $3 \times 10^{4}$\,cm$^2$\,s$^{-1}$ has been found for this target. Interestingly, this value is similar to the viscosities obtained from models aiming at reproducing the internal rotation of the Sun \cite[e.g.][]{rue96,spa10} and to viscosities deduced from observations of the spin-down of stars in open clusters \cite[e.g.][]{den10_spin}. Does this mean that the same efficiency of internal angular momentum transport is required during the main-sequence and the red-giant phase? Recalling that KIC~8366239 is a 1.5\,M$_{\odot}$ early red giant, does this mean that the same efficiency is required for stars with different masses, and in particular for stars with a convective (such as KIC~8366239) or a radiative core during the main sequence?

As a first step to study how the efficiency of the required additional transport process changes with stellar properties, we try to characterize the additional viscosity needed for the low-mass red giant KIC~7341231. This is done by using the values of the core rotation rate $\Omega_{\rm c} = 710 \pm 51$\,nHz and of the upper limit on the surface rotation rate $\Omega_{\rm s} < 150 \pm 19$\,nHz determined by \cite{deh12} from rotational splittings of mixed modes. Following the results obtained for KIC~8366239, we first compute a small grid of rotating models for KIC~7341231 with an additional viscosity fixed to $\nu_{\rm add} = 3 \times 10^{4}$\,cm$^2$\,s$^{-1}$. These first models are computed for different initial rotation velocities on the ZAMS and without braking of the stellar surface by magnetized winds. The latter assumption is of course not realistic in view of observations of surface rotation rates for main-sequence solar-type stars, but is especially interesting to determine whether the post-main-sequence transport of angular momentum can be constrained from asteroseismic measurements of red giants independently of the past history of the star. This is particularly important in view of the uncertainties about the exact modelling and efficiency of wind braking especially for stars that are different from the Sun \citep[see for instance][]{van16}. This point is addressed in Sect.~\ref{brake} by comparing the results obtained with these first models without wind braking (which represent an extreme case of an inefficient surface braking) with results obtained from models computed with wind braking. Note that the neglect of wind braking leads to low values of initial velocities on the ZAMS to correctly reproduce the rotation rates deduced from asteroseismic measurements of KIC~7341231.

\begin{figure}[htb!]
\resizebox{\hsize}{!}{\includegraphics{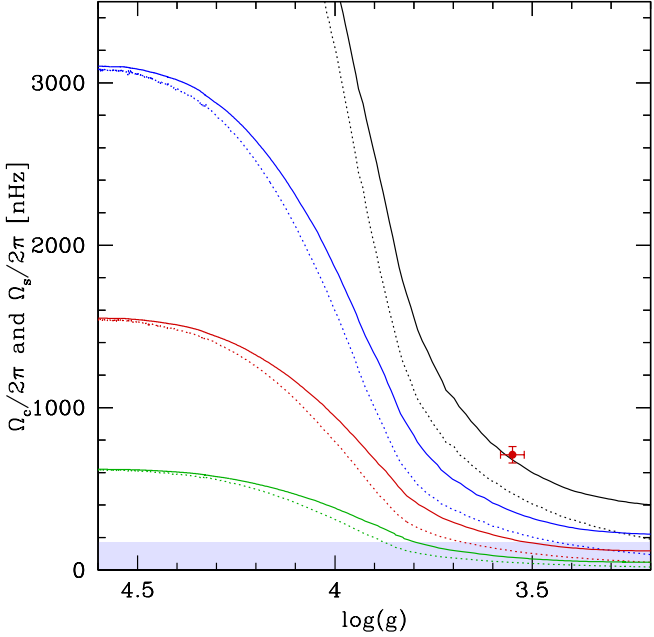}}
 \caption{Core (continuous lines) and surface (dotted lines) rotation rates as a function of gravity for rotating models of the red giant KIC~7341231 computed with an additional viscosity $\nu_{\rm add} = 3 \times 10^{4}$\,cm$^2$\,s$^{-1}$. Black, blue, red and green lines indicate models computed with an initial velocity on the ZAMS of 20, 10, 5, and 1\,km\,s$^{-1}$, respectively. The red dot indicates the values of the core rotation rate and gravity determined for KIC~7341231, while the blue region indicates the upper limit for its surface rotation rate \citep{deh12}.}
  \label{omegacs_Otto_logg_Dadd3e4}
\end{figure}

The evolution of the core and surface rotation rates for rotating models of the red giant KIC~7341231 computed with an additional viscosity of $3 \times 10^{4}$\,cm$^2$\,s$^{-1}$ is shown in Fig.~\ref{omegacs_Otto_logg_Dadd3e4}. Continuous and dotted lines correspond to core and surface rotation rates, respectively. The evolution is shown from the ZAMS to the early red-giant phase as a function of gravity. The shaded blue region corresponds to the upper limit of $169$\,nHz on the surface rotation rate of KIC~7341231. The red dot indicates the core rotation rate of $710 \pm 51$\,nHz and the asteroseismic value of gravity $\log g =3.55 \pm 0.03$ determined for KIC~7341231 \citep[see][for more details]{deh12}. Colors correspond to the initial velocities of the models on the ZAMS: black, blue, red and green lines indicate ZAMS velocities of 20, 10, 5, and 1\,km\,s$^{-1}$, respectively. Figure~\ref{omegacs_Otto_logg_Dadd3e4} shows that none of these models is able to simultaneously reproduce the asteroseismic constraints on the core and surface rotation rates available for KIC~7341231. Models with low initial rotation velocities are compatible with the upper limit on the surface rotation rate, but predict too low core rotation rates. Similarly, a model with a higher initial velocity on the ZAMS of 20\,km\,s$^{-1}$ correctly reproduces the core rotation rate of KIC~7341231, but predicts a too high surface rotation rate. We thus conclude that the viscosity $\nu_{\rm add} = 3 \times 10^{4}$\,cm$^2$\,s$^{-1}$ leads to a too efficient transport of angular momentum in the radiative interior, which is not compatible with the lower limit on the degree of radial differential rotation deduced from asteroseismic data of KIC~7341231. The mean viscosity corresponding to the undetermined transport process is thus found to be lower for KIC~7341231 than for KIC~8366239.

\subsection{Determination of the additional viscosity}
\label{nobrake}

We have just seen that the mean value of additional viscosity required in the case of the more massive red giant KIC~8366239 leads to a too efficient coupling to correctly reproduce the asteroseismic constraints available for KIC~7341231. We now investigate which constraints can be brought on the efficiency of the additional transport mechanism for this star. 

While the determination of the additional viscosity is performed here for the specific case of KIC~7341231, we note that such a study is more general. It illustrates the constraints that can be obtained about the internal transport of angular momentum in the case of red giants, for which a value of the core rotation rate but only an upper limit on the surface rotation are deduced from asteroseismic measurements. We recall here that obtaining only an upper limit on the surface rotation rate from asteroseismic data is common for stars on the red giant branch because it is difficult to evaluate the contribution from the central layers to the envelope kernel \citep[see Sect.~5 of][]{deh12}.

\begin{figure}[htb!]
\resizebox{\hsize}{!}{\includegraphics{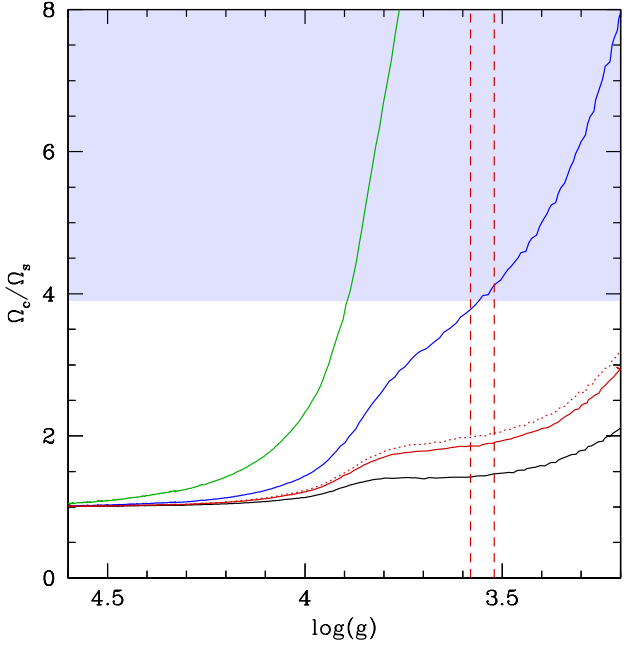}}
 \caption{Ratio of core to surface rotation rate as a function of gravity for rotating models of KIC~7341231. Black, red, blue and green lines correspond to models computed with an additional viscosity of $3 \times 10^{4}$, $2 \times 10^{4}$, $1 \times 10^{4}$ and $3 \times 10^{3}$\,cm$^2$\,s$^{-1}$, and initial velocities on the ZAMS of 20, 15, 8, and 2\,km\,s$^{-1}$, respectively. The red dotted line corresponds to a model with an additional viscosity of $2 \times 10^{4}$\,cm$^2$\,s$^{-1}$ and an initial velocity on the ZAMS of 10\,km\,s$^{-1}$. Red dashed vertical lines indicate the values of the gravity, while the blue region indicates the lower limit of core-to-surface rotation rate determined for KIC~7341231 \citep{deh12}.}
  \label{rap_omegasc_Daddmax}
\end{figure}

\begin{figure}[htb!]
\resizebox{\hsize}{!}{\includegraphics{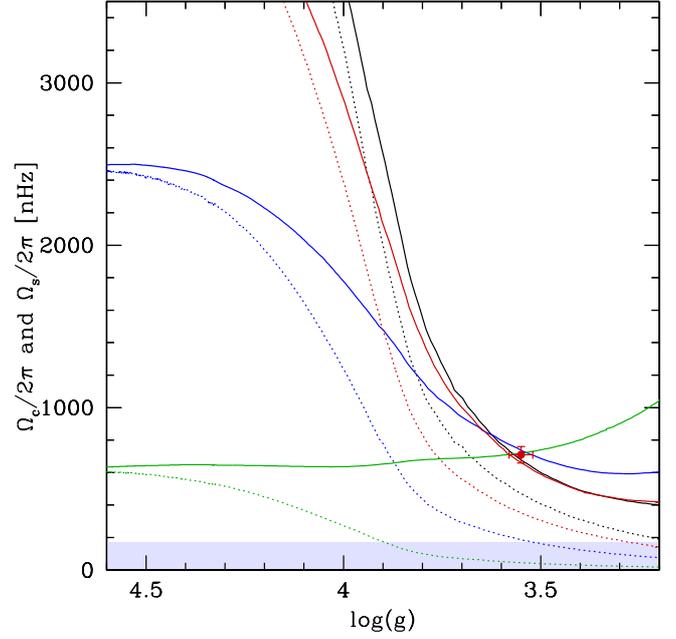}}
 \caption{Same as Fig.~\ref{omegacs_Otto_logg_Dadd3e4} but for models of KIC~7341231 computed with different values of the additional viscosity. Black, red, blue and green lines correspond to an additional viscosity of $3 \times 10^{4}$, $2 \times 10^{4}$, $1 \times 10^{4}$ and $3 \times 10^{3}$\,cm$^2$\,s$^{-1}$, and initial velocities on the ZAMS of 20, 15, 8, and 2\,km\,s$^{-1}$, respectively.}
  \label{omegacs_Otto_logg_Daddmax}
\end{figure}

Rotating models that correctly reproduce the global properties of KIC~7341231 are computed for various values of the additional viscosity $\nu_{\rm add}$ and different initial velocities on the ZAMS. For a given value of the additional viscosity, an initial velocity on the ZAMS is then determined to predict a core rotation rate that is compatible with the one deduced from asteroseismic measurements. Figure~\ref{rap_omegasc_Daddmax} shows the change in the ratio between the core and the surface rotation rates as a function of gravity for these models. As expected, the contrast between the rotation in the core and at the stellar surface decreases when the value of the additional viscosity increases, since the internal transport of angular momentum becomes more and more efficient. The ratio of the core to the surface rotation rate is found to be almost insensitive to the adopted initial velocity on the ZAMS. This can be seen by comparing the continuous and dotted red lines in Fig.~\ref{rap_omegasc_Daddmax}, which correspond to models computed with the same additional viscosity of $2 \times 10^{4}$\,cm$^2$\,s$^{-1}$ and an initial velocity of the ZAMS of 15 and 10\,km\,s$^{-1}$, respectively. This illustrates that the additional viscosity dominates the transport of angular momentum. Of course, both models can be distinguished by using the additional constraint on the core rotation rate: in Fig.~\ref{rap_omegasc_Daddmax}, all continuous lines indicate models that correctly reproduce the observed core rotation rate (see Fig.~\ref{omegacs_Otto_logg_Daddmax}), while the model shown by the dotted line predicts a too low rotation rate for KIC~7341231 due to its lower initial velocity.

As shown in Fig.~\ref{rap_omegasc_Daddmax}, obtaining an upper limit on the surface rotation rate from asteroseismic data in addition to the core rotation rate enables a precise determination of the maximum mean efficiency allowed for the additional transport mechanism in KIC~7341231. Combining the lower value of the core rotation rate ($659$\,nHz) with the upper limit on the surface rotation rate of 169\,nHz leads to a minimal contrast between the angular velocity in the core and at the surface of about 3.9 for KIC~7341231 (blue shaded region in Fig.~\ref{rap_omegasc_Daddmax}). Rotating models computed with too high values of the additional viscosity can then be discarded using this limit. From Fig.~\ref{rap_omegasc_Daddmax}, one sees that only models with $\nu_{\rm add}$ lower than $1 \times 10^{4}$\,cm$^2$\,s$^{-1}$ are able to reproduce this observational constraint. This is also illustrated in Fig.~\ref{omegacs_Otto_logg_Daddmax}, which shows the evolution of the core and surface rotation rates as a function of gravity for the models indicated by continuous lines in Fig.~\ref{rap_omegasc_Daddmax}. Models with high values of the additional viscosity (for instance the black and red lines corresponding to $\nu_{\rm add}=3 \times 10^{4}$ and $2 \times 10^{4}$\,cm$^2$\,s$^{-1}$, respectively) are then characterized by too high surface rotation rates compared to asteroseismic constraints.

While having an upper limit on the surface rotation rate is valuable for determining the maximum efficiency of the additional process for the internal transport of angular momentum, this does not provide any constraint on the minimum efficiency needed for this mechanism. According to Figs.~\ref{rap_omegasc_Daddmax} and \ref{omegacs_Otto_logg_Daddmax}, all models with $\nu_{\rm add}$ lower than $1 \times 10^{4}$\,cm$^2$\,s$^{-1}$ are compatible with the asteroseismic constraints on the core and surface rotation rates. It seems then difficult to obtain a lower limit on the additional viscosity required for KIC~7341231 based solely on these constraints. To reach the value of the core rotation rate obtained for KIC~7341231, one however sees in Fig.~\ref{omegacs_Otto_logg_Daddmax} that the initial velocity on the ZAMS has to be significantly decreased when the additional viscosity decreases. In addition to the asteroseismic constraints on the core and surface rotation rates, one can then add a constraint on the minimum value of the initial rotation velocity on the ZAMS that seems reasonable according to observations of rotation periods in young open clusters. In the present study, we then assume that the initial rotation velocity of KIC~7341231 on the ZAMS was not lower than 1\,km\,s$^{-1}$. For the 0.84\,M$_{\odot}$ star KIC~7341231, this initial rotation velocity on the ZAMS of 1\,km\,s$^{-1}$ translates into a surface angular velocity of about 0.7\,$\Omega_{\odot}$ (rotation period of 36 days) on the ZAMS. According to observations of rotation periods in young open clusters, this corresponds to a conservative lower limit. Using for instance the sample of stars in young open clusters studied by \cite{gal15}, one indeed notes that no stars are observed on the ZAMS with surface angular velocities lower than 1\,$\Omega_{\odot}$ in the mass range 0.7 -- 0.9\,M$_{\odot}$, while a mean value for the 25th rotational percentile of about 4.5\,$\Omega_{\odot}$ is observed on the ZAMS for this mass range.

\begin{figure}[htb!]
\resizebox{\hsize}{!}{\includegraphics{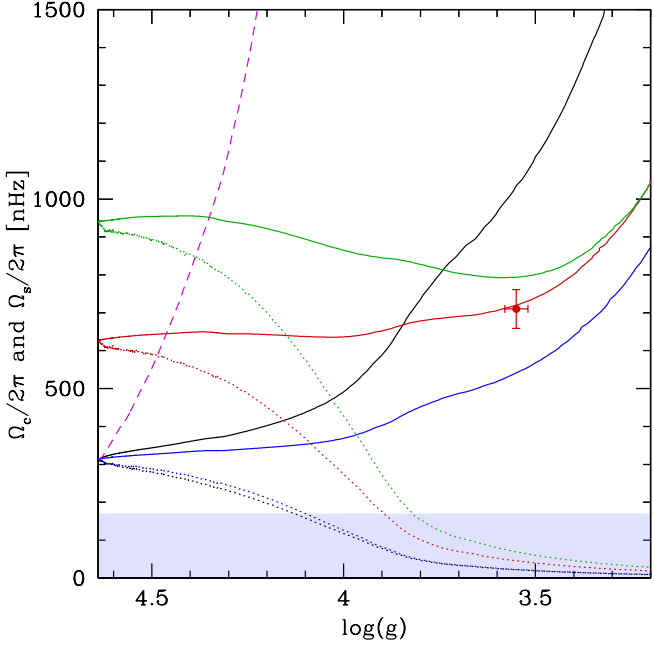}}
 \caption{Same as Fig.~\ref{omegacs_Otto_logg_Dadd3e4} but for models of KIC~7341231 computed with low initial rotation rates. The dashed line indicates the core rotation rate for a model computed with shellular rotation only (i.e. $\nu_{\rm add} = 0$) and an initial velocity on the ZAMS of 1\,km\,s$^{-1}$. The black and blue lines indicate models computed with an initial velocity on the ZAMS of 1\,km\,s$^{-1}$ and with an additional viscosity of $1 \times 10^{3}$ and $2 \times 10^{3}$\,cm$^2$\,s$^{-1}$, respectively. The red line corresponds to a model with an initial velocity on the ZAMS of 2\,km\,s$^{-1}$ and an additional viscosity of $3 \times 10^{3}$\,cm$^2$\,s$^{-1}$. The green line indicates a model with an initial velocity on the ZAMS of 3\,km\,s$^{-1}$ and an additional viscosity of $4 \times 10^{3}$\,cm$^2$\,s$^{-1}$.}
  \label{omegacs_Otto_logg_Daddmin_v2}
\end{figure}

The variation of the core and surface rotation rates as a function of the surface gravity is shown in Fig.~\ref{omegacs_Otto_logg_Daddmin_v2} for models of KIC~7341231 that are slow rotators on the ZAMS. The dashed magenta line illustrates the evolution of the core rotation rate for a model that does not take into account an additional process for the internal transport of angular momentum. Despite a very low initial velocity on the ZAMS of 1\,km\,s$^{-1}$, such a model predicts a very rapidly rotating core during the red-giant phase, in contradiction with the core rotation rate deduced from asteroseismic data \cite[see also][]{cei13}. As explained above, models with an initial velocity on the ZAMS of 1\,km\,s$^{-1}$ are then used to estimate the lower limit required for the efficiency of this process. The continuous black and blue lines in Fig.~\ref{omegacs_Otto_logg_Daddmin_v2} correspond to such models computed with an additional viscosity of $1 \times 10^{3}$ and $2 \times 10^{3}$\,cm$^2$\,s$^{-1}$, respectively. The model with $\nu_{\rm add}=1 \times 10^{3}$\,cm$^2$\,s$^{-1}$ (black line) predicts a slightly higher rotation rate in the core compared to the asteroseismic value, while the model with $\nu_{\rm add}=2 \times 10^{3}$\,cm$^2$\,s$^{-1}$ (blue line) leads to a too low core rotation rate. Figure~\ref{omegacs_Otto_logg_Daddmin_v2} also shows that the required additional viscosity rapidly increases when the initial rotation on the ZAMS increases. The red and green lines indeed indicate that an additional viscosity of $3 \times 10^{3}$\,cm$^2$\,s$^{-1}$ and a value slightly lower than $4 \times 10^{3}$\,cm$^2$\,s$^{-1}$ is needed for an initial velocity of 2 and 3\,km\,s$^{-1}$, respectively. We thus conclude that a value of $\nu_{\rm add}=1 \times 10^{3}$\,cm$^2$\,s$^{-1}$ can be safely used as an estimate for the lower limit on the efficiency of the unknown additional mechanism for angular momentum transport. Combining this with the upper limit on the transport efficiency deduced solely from asteroseismic measurements, one concludes that the mean viscosity that characterizes this undetermined transport process can be constrained to $\nu_{\rm add} = 1 \times 10^{3}$ -- $1 \times 10^{4} $\,cm$^2$\,s$^{-1}$ in the case of the red giant KIC~7341231.

\section{Impact of the braking of the stellar surface by magnetized winds}
\label{brake}

The above results about the efficiency of the additional process for internal angular momentum transport have been obtained by computing rotating models of KIC~7341231 that neglect a possible braking of the stellar surface by magnetized winds. Stellar models of the low-mass red giant KIC~7341231 exhibit a convective envelope during the whole main-sequence evolution, which could generate magnetic fields through a dynamo process and hence a braking of the surface by magnetized winds. By increasing the radial differential rotation, one might expect that this braking changes the values for the additional viscosity required for this target. To investigate this point, rotating models of KIC~7341231 are computed by including a braking of the stellar surface by magnetized winds. For this purpose, we adopt the braking law of \cite{kri97}, which expresses the related torque on the stellar surface by the following relation:
\begin{eqnarray}
 \frac{{\rm d} J}{{\rm d}t} = \left\{
\begin{array}{l l }
-K \Omega^3 \left({\displaystyle \frac{R}{R_\odot}} \right)^{1/2} 
\left({\displaystyle \frac{M}{M_\odot} }\right)^{-1/2} & 
(\Omega \leq \Omega_{\rm sat}) \\
 & \\
-K \Omega \, {\Omega^2}_{\rm sat} \left({\displaystyle \frac{R}{R_\odot}} \right)^{1/2} 
\left({\displaystyle \frac{M}{M_\odot} }\right)^{-1/2}
& (\Omega > \Omega_{\rm sat}) \, .  
 \end{array}   \right.
\label{eq_braking}
\end{eqnarray}
$\Omega_{\rm sat}$ is introduced to account for the saturation of the magnetic field generation; this critical value is fixed to 8\,$\Omega_{\odot}$ in the present computations. The braking constant $K$ is calibrated so that rotating solar models reproduce the solar surface rotational velocity after $4.57$\,Gyr ($K_{\odot}=2.7 \times 10^{47}$\,cm$^2$\,g\,s).

\begin{figure}[htb!]
\resizebox{\hsize}{!}{\includegraphics{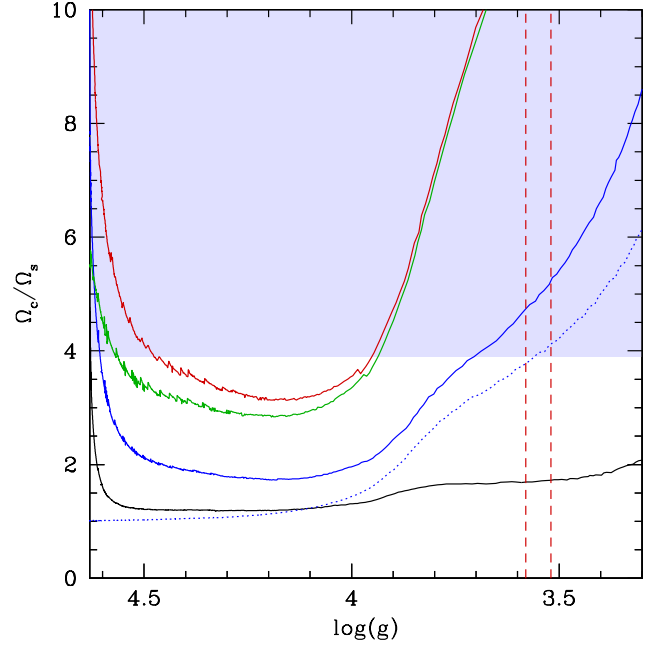}}
 \caption{Ratio of core-to-surface rotation rate as a function of gravity for models that include magnetic braking of the stellar surface (with a solar-calibrated value for the braking constant). Black, blue and red continuous lines correspond to an initial velocity on the ZAMS of 50\,km\,s$^{-1}$ and an additional viscosity of $3 \times 10^{4}$, $1 \times 10^{4}$ and $4 \times 10^{3}$\,cm$^2$\,s$^{-1}$, respectively. The green line corresponds to a similar model with an initial velocity on the ZAMS of 10\,km\,s$^{-1}$ and $\nu_{\rm add}=4 \times 10^{3}$\,cm$^2$\,s$^{-1}$. The dotted blue line corresponds to a model without surface magnetic braking, with $\nu_{\rm add}=1 \times 10^{4}$\,cm$^2$\,s$^{-1}$ and an initial velocity on the ZAMS of 8\,km\,s$^{-1}$. Red dashed vertical lines indicate the values of the gravity of KIC~7341231, while the blue region indicates the lower limit of core-to-surface rotation rate determined for KIC~7341231.}
  \label{rap_omegasc_frein_tot}
\end{figure}

Figure~\ref{rap_omegasc_frein_tot} shows the evolution of the ratio between the angular velocity in the core and at the surface for rotating models of KIC~7341231 that take into account the braking by magnetized winds as expressed by Eq.~\ref{eq_braking}. As discussed in the preceding section, such a plot is interesting to first estimate an upper limit on the efficiency of the additional transport process. Including magnetic braking does not change the conclusion about the fact that the additional viscosity of $3 \times 10^{4}$\,cm$^2$\,s$^{-1}$ determined for the more massive red giant KIC~8366239 (black line in Fig.~\ref{rap_omegasc_frein_tot}) is too high to correctly reproduce the lower limit on the degree of radial differential rotation obtained for KIC~7341231. One however notes that the inclusion of surface magnetic braking leads to a slight increase in the ratio of the core-to-surface rotation rates for models computed with the same value of the additional viscosity. This can be seen by comparing models with $\nu_{\rm add}=1 \times 10^{4}$\,cm$^2$\,s$^{-1}$ (blue lines in Fig.~\ref{rap_omegasc_frein_tot}), which are computed with (continuous line) and without surface braking (dotted line). At first sight, this suggests that the braking of the stellar surface by magnetized winds could result in an increase of the upper limit on the efficiency of the additional transport mechanism. By decreasing the angular velocity of the convective envelope, magnetized winds increase the radial differential rotation. This increase must then be compensated for by a more efficient transport of angular momentum to reach the same ratio of core-to-surface rotation rates. Figure~\ref{rap_omegasc_frein_tot} also illustrates that the rotational properties of models including surface magnetic braking rapidly converge when the initial rotation on the ZAMS increases. This can be seen by comparing models with the same value of $\nu_{\rm add}=4 \times 10^{3}$\,cm$^2$\,s$^{-1}$ but different initial velocities on the ZAMS (red and green lines in Fig.~\ref{rap_omegasc_frein_tot}).

\begin{figure}[htb!]
\resizebox{\hsize}{!}{\includegraphics{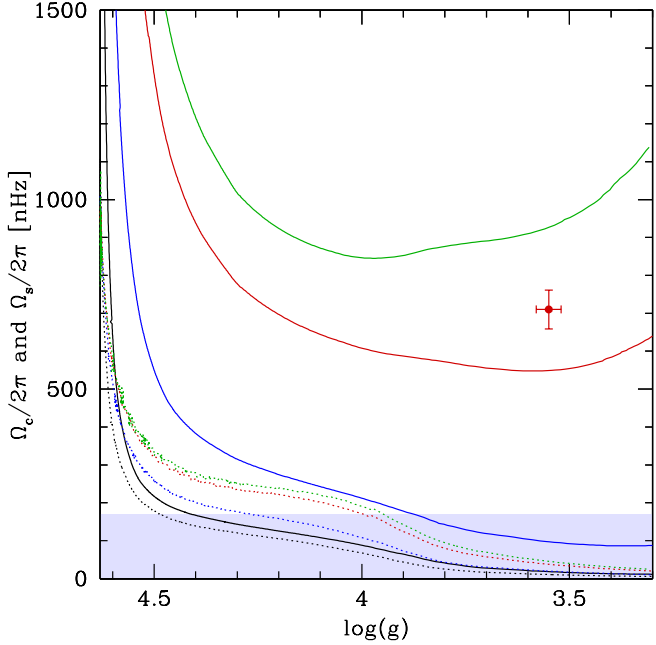}}
 \caption{Core (continuous lines) and surface (dotted lines) rotation rates as a function of gravity for models with an initial velocity on the ZAMS of 50\,km\,s$^{-1}$, which account for surface magnetic braking. Black, blue, red and green lines indicate models with a solar-calibrated braking and an additional viscosity of $3 \times 10^{4}$, $1 \times 10^{4}$, $4 \times 10^{3}$ and $3 \times 10^{3}$\,cm$^2$\,s$^{-1}$, respectively. The red dot indicates the values of the core rotation rate and gravity determined for KIC~7341231, while the blue region indicates the upper limit for its surface rotation rate.}
  \label{omegacs_vs_logg_frein_vi50}
\end{figure}

While Fig.~\ref{rap_omegasc_frein_tot} suggests that the upper limit on the additional viscosity needed for KIC~7341231 might be slightly higher when surface braking is accounted for, we compare in Fig.~\ref{omegacs_vs_logg_frein_vi50} the predictions of these models with the value deduced for the core rotation rate. We first note that in addition to predicting a too low degree of radial differential rotation, a viscosity of $3 \times 10^{4}$\,cm$^2$\,s$^{-1}$ also leads to a too low core rotation rate (black line in Fig.~\ref{omegacs_vs_logg_frein_vi50}). Interestingly, this is also the case for the model with $\nu_{\rm add}=1 \times 10^{4}$\,cm$^2$\,s$^{-1}$ (blue line in Fig.~\ref{omegacs_vs_logg_frein_vi50}); even the model with $\nu_{\rm add}=4 \times 10^{3}$\,cm$^2$\,s$^{-1}$ (red line in Fig.~\ref{omegacs_vs_logg_frein_vi50}) exhibits a slightly lower core rotation rate than observed. Instead of increasing the value of the upper limit on the efficiency of the additional transport process, the inclusion of surface magnetic braking thus leads to a lower value for this limit (lower than $4 \times 10^{3}$\,cm$^2$\,s$^{-1}$ for a solar-calibrated braking constant). This is due to the simultaneous efficient internal transport of angular momentum through the additional viscosity and loss of angular momentum at the stellar surface by magnetized winds. Both effects result in a significant decrease of the angular momentum in the central stellar layers and then in a low core rotation rate at the base of the red giant branch, which is incompatible with the value deduced from asteroseismic data for KIC~7341231. This conclusion is in good agreement with the results of \cite{tay13} obtained by using the simple assumption of solid-body rotation of the whole star.

\begin{figure}[htb!]
\resizebox{\hsize}{!}{\includegraphics{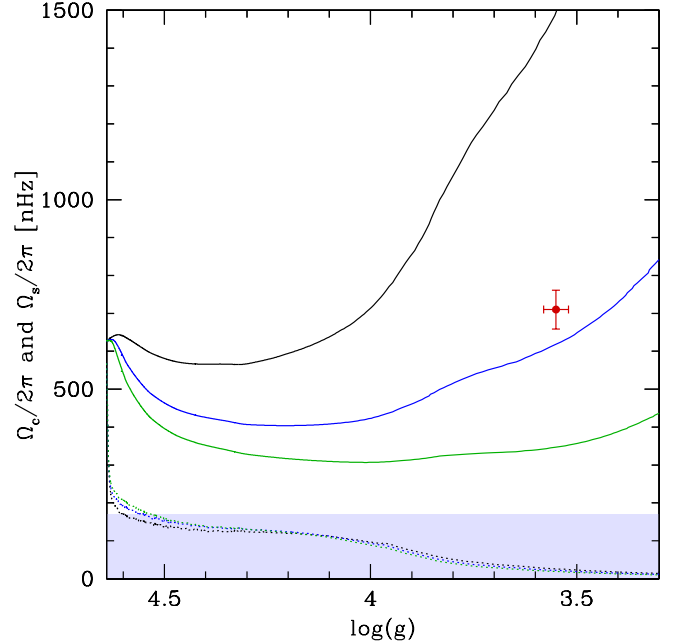}}
 \caption{Same as Fig.~\ref{omegacs_vs_logg_frein_vi50} but for models with an initial velocity on the ZAMS of 2\,km\,s$^{-1}$. The black, blue and green lines indicate models computed with a solar-calibrated braking of the stellar surface during the main sequence and an additional viscosity of $1 \times 10^{3}$, $2 \times 10^{3}$ and $3 \times 10^{3}$\,cm$^2$\,s$^{-1}$, respectively.}
  \label{omegacs_vs_logg_frein_vi2}
\end{figure}

Taking into account a braking of the stellar surface by magnetized winds has only a limited impact on the minimum value of the additional viscosity needed to reproduce the core and surface rotation rates of KIC~7341231. This is shown in Fig.~\ref{omegacs_vs_logg_frein_vi2} for models computed with the same low initial rotation velocity of 2\,km\,s$^{-1}$ on the ZAMS. Models computed with a surface braking (using a solar-calibrated value for the braking constant) indicate that a viscosity of about $2 \times 10^{3}$\,cm$^2$\,s$^{-1}$ is obtained (blue line in Fig.~\ref{omegacs_vs_logg_frein_vi2}), while models without braking lead to a slightly higher value of $3 \times 10^{3}$\,cm$^2$\,s$^{-1}$ (red line in Fig.~\ref{omegacs_vs_logg_frein_vi2}). The slight decrease in the additional viscosity that is needed to reproduce the core rotation rate of KIC~7341231 for models that include surface magnetic braking is also a direct consequence of the loss of angular momentum by magnetized winds.

\section{Impact of the internal transport of angular momentum during the main sequence}

In Sect.~\ref{nobrake}, we showed that the mean efficiency of the missing transport mechanism for angular momentum can be constrained to the range $\nu_{\rm add} = 1 \times 10^{3}$ -- $1 \times 10^{4} $\,cm$^2$\,s$^{-1}$ using the asteroseismic values of the core rotation rate and the upper limit on the surface rotation rate obtained for KIC~7341231. This range of viscosity is obtained by neglecting the braking of the stellar surface by magnetized winds. The effects of such an assumption have been studied in Sect.~\ref{brake}. Interestingly, the range in $\nu_{\rm add}$ is significantly reduced when a braking of the stellar surface is taken into account, with $\nu_{\rm add} = 1 \times 10^{3}$ -- $4 \times 10^{3} $\,cm$^2$\,s$^{-1}$ for a solar-calibrated values of the braking constant. 

It is important to underline here that all these results have been obtained by assuming that the unknown transport process is at work from the beginning of the evolution of the star and in particular during the main sequence. Such an assumption seems reasonable in view of the need for an additional efficient angular momentum transport in the solar case (see discussion in Sect.~\ref{intro}). In this sense, the above values for the additional viscosity correspond to a mean efficiency during the whole evolution of the star. However, the main aim of introducing an additional viscosity corresponding to the unknown transport mechanism in the equation describing the angular momentum transport is to determine how asteroseismic observations of red giants can constrain such a process. In particular, it is crucial to investigate whether such asteroseismic measurements can precisely constrain the efficiency of the transport of angular momentum after the main sequence, or whether the efficiency deduced from asteroseismic data of red giants is dominated by the earlier evolution on the main sequence.

\begin{figure}[htb!]
\resizebox{\hsize}{!}{\includegraphics{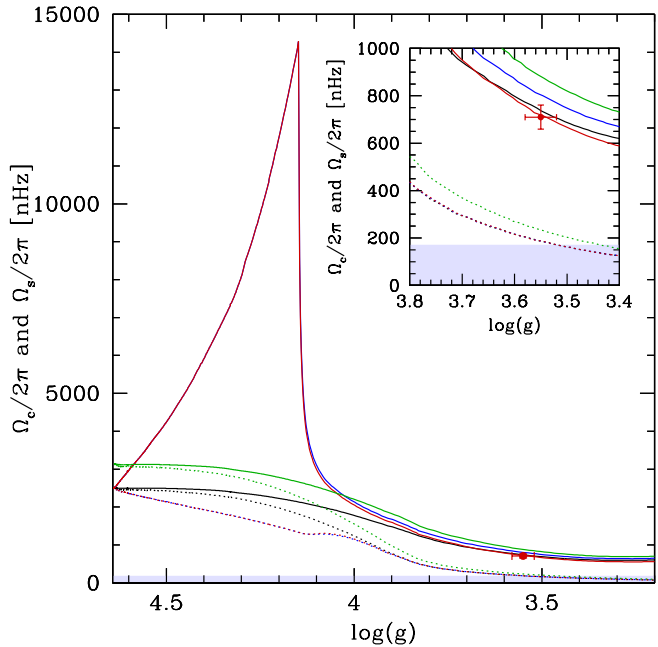}}
 \caption{Core (continuous lines) and surface (dotted lines) rotation rates for models computed with and without an additional viscosity during the main sequence. These models do not include a braking of the surface by magnetized winds. The black and green lines indicate models with an additional viscosity of $1 \times 10^{4}$\,cm$^2$\,s$^{-1}$ during the whole evolution and an initial velocity on the ZAMS of 8 and 10\,km\,s$^{-1}$, respectively. The blue and red lines correspond to models without an additional viscosity during the main sequence, with an initial velocity on the ZAMS of 8\,km\,s$^{-1}$, and an additional viscosity during the post-main sequence of $1 \times 10^{4}$ and $1.1 \times 10^{4}$\,cm$^2$\,s$^{-1}$, respectively.}
  \label{omegacs_logg_Daddmax_postMS}
\end{figure}

For this purpose, new models of KIC~7341231 are computed without an additional viscosity during the evolution on the main sequence; the additional viscosity is then only introduced during the post-main-sequence evolution of the star. These models are first computed without braking of the stellar surface by magnetized winds. The evolution of the core and surface rotation rates of such models are shown in Fig.~\ref{omegacs_logg_Daddmax_postMS} for an initial rotation velocity on the ZAMS of 8\,km\,s$^{-1}$ (blue and red lines in Fig.~\ref{omegacs_logg_Daddmax_postMS}). The low efficiency of the transport of angular momentum by meridional currents and the shear instability in radiative zones leads to a large increase in the core rotation rate during the main sequence for models computed without an additional viscosity. This results in a huge increase of the radial differential rotation at the end of the main sequence (compare for instance the black and blue continuous lines in Fig.~\ref{omegacs_logg_Daddmax_postMS}). At the beginning of the post-main-sequence evolution, the additional viscosity is taken into account, which explains the rapid decrease in the core rotation rate for models without an additional viscosity on the main sequence (at a $\log g$ of about 4.1 in Fig.~\ref{omegacs_logg_Daddmax_postMS}), and the slight increase in their surface rotation rate. Despite the high value of the core rotation rate reached at the end of the main sequence by these models, we find that the value of the additional viscosity needed during the post-main sequence is very similar to the one deduced by assuming the unknown transport mechanism to be at work during the whole evolution of the star. This is clearly seen in the zoom shown in Fig.~\ref{omegacs_logg_Daddmax_postMS}. The black and blue lines correspond to exactly the same models (same initial velocity of 8\,km\,s$^{-1}$ and same $\nu_{\rm add} = 1 \times 10^{4}$\,cm$^2$\,s$^{-1}$) computed with and without the additional viscosity during the main sequence, respectively. The fact that the additional viscosity is not accounted for during the main sequence leads only to a very small increase in the core rotation rate predicted for KIC~7341231 while it almost fails to affect the surface rotation rate. To correctly reproduce the precise asteroseismic determination of the core rotation rate of KIC~7341231, a maximum value of the additional viscosity of $1.1 \times 10^{4}$\,cm$^2$\,s$^{-1}$ (red line in Fig.~\ref{omegacs_logg_Daddmax_postMS}) is found for models without an additional viscosity on the main sequence (instead of $1 \times 10^{4}$\,cm$^2$\,s$^{-1}$ when the additional viscosity is included during the whole evolution of the star). We thus see that the efficiency of the internal transport of angular momentum during the main sequence has a negligible impact on the value of the additional viscosity needed to account for the asteroseismic determination of the rotational properties of KIC~7341231.

\begin{figure}[htb!]
\resizebox{\hsize}{!}{\includegraphics{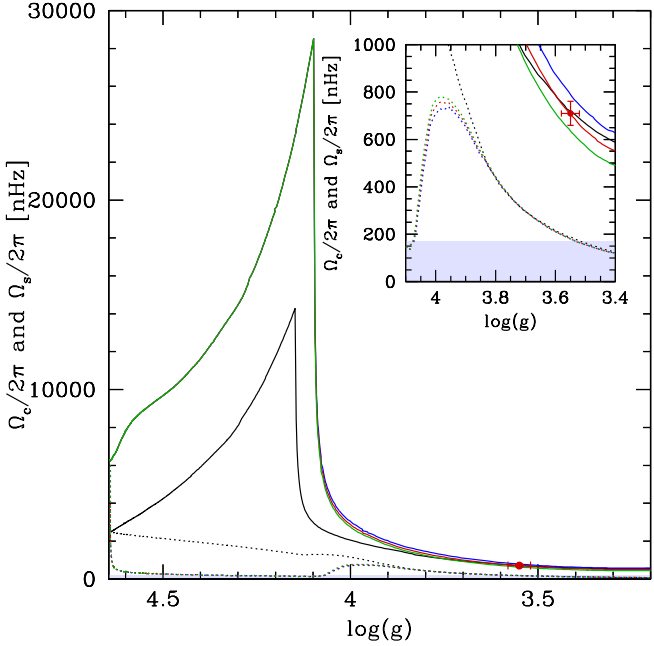}}
 \caption{Same as Fig.~\ref{omegacs_logg_Daddmax_postMS} for models computed without additional viscosity during the main sequence. The black line corresponds to a model with an initial velocity on the ZAMS of 8\,km\,s$^{-1}$, an additional viscosity during the post-main sequence of $1.1 \times 10^{4}$\,cm$^2$\,s$^{-1}$, and without braking of the surface by magnetized winds. Blue, red and green lines indicate models that include a solar-calibrated braking of the stellar surface and an additional viscosity during the post-main sequence of $1.1 \times 10^{4}$, $1.2 \times 10^{4}$ and $1.3 \times 10^{4}$\,cm$^2$\,s$^{-1}$, respectively. They are computed for an initial velocity of 20\,km\,s$^{-1}$ on the ZAMS.}
  \label{omegacs_logg_Daddmax_postMS_frein}
\end{figure}

Finally, rotating models of KIC~7341231 without an additional viscosity during the main sequence, but with a solar-calibrated braking of the surface by magnetized winds are computed. The low efficiency of the transport of angular momentum by meridional circulation and shear instability together with the braking of the stellar surface leads to a high degree of radial differential rotation on the main sequence for these models (see Fig.~\ref{omegacs_logg_Daddmax_postMS_frein}). This is of course in contradiction with the approximately flat rotation profile of the Sun, but this corresponds to an extreme case, which is especially interesting to test the sensitivity of the mean efficiency for the internal transport of angular momentum deduced from asteroseismic measurements of red giants on the past rotational history of the star. The core and surface rotation rates of these models are shown in Fig.~\ref{omegacs_logg_Daddmax_postMS_frein} for different values of the additional viscosity, which is taken into account only during the post-main-sequence evolution. While the contrast between the rotation velocity in the core and at the surface is much higher at the end of the main sequence for models computed with a surface braking than for models computed without braking, we find that the value obtained for the additional viscosity is very similar in both cases. Only a slight increase of $\nu_{\rm add}$ to $1.2$ instead of $1.1 \times 10^{4}$\,cm$^2$\,s$^{-1}$ is needed to reach the same values for the core and surface rotation rate of KIC~7341231 when a surface braking is taken into account (black and red lines in Fig.~\ref{omegacs_logg_Daddmax_postMS_frein}). As shown with the continuous green line in Fig.~\ref{omegacs_logg_Daddmax_postMS_frein}, a maximum value of $\nu_{\rm add}=1.3 \times 10^{4}$\,cm$^2$\,s$^{-1}$ is found to correctly reproduce the core rotation rate of KIC~7341231. Using models computed with different assumptions for the modelling of rotational effects, we thus see that a similar value of about $1 \times 10^{4}$\,cm$^2$\,s$^{-1}$ is found for the upper limit on the efficiency of the unknown mechanism for the transport of angular momentum. This shows that the asteroseismic constraints on the rotation rates of red giants are able to provide us with strong constraints on the efficiency of this internal transport of angular momentum during the post-main-sequence phase, without being sensitive to the assumptions made for such a transport during the main sequence.

\begin{figure}[htb!]
\resizebox{\hsize}{!}{\includegraphics{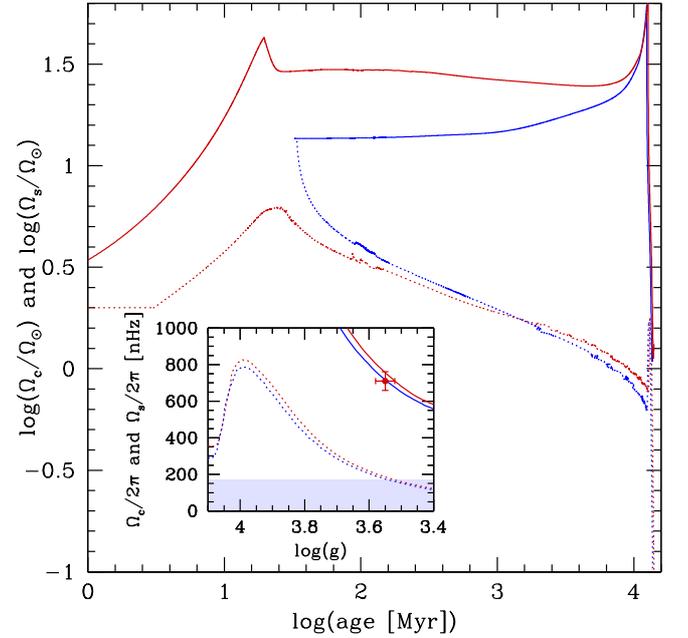}}
 \caption{Core (continuous lines) and surface (dotted lines) rotation rates as a function of age for a rotating model of KIC~7341231 starting on the ZAMS (blue lines) and a model that includes the pre-main-sequence evolution (red lines). The blue lines correspond to a model with solid-body rotation on the ZAMS and an initial rotation velocity of 20\,km\,s$^{-1}$. The red lines correspond to a model with an initial angular velocity of 2\,$\Omega_{\odot}$ and a disc lifetime of 3\,Myr. Both models are computed with hydrodynamic processes only before the post-main sequence, and with $\nu_{\rm add}=1.2 \times 10^{4}$\,cm$^2$\,s$^{-1}$ during the post-main sequence. A zoom on the variation of the core and surface rotation rates as a function of $\log g$ is shown in the inset for these models. The red dot corresponds to the values of the core rotation rate and gravity determined for KIC~7341231, while the blue region indicates the upper limit for its surface rotation rate.}
  \label{omegas_vs_t_fig}
\end{figure}

This conclusion is also valid for the uncertainties related to the treatment of rotational effects during the pre-main-sequence. This is illustrated in Fig.~\ref{omegas_vs_t_fig}, where a model of KIC~7341231 computed with shellular rotation during the pre-main sequence is compared with a model computed by assuming solid-body rotation on the ZAMS. The model that includes the pre-main sequence phase has an initial angular velocity of 2\,$\Omega_{\odot}$ and a disc lifetime of 3\,Myr. The surface angular velocity is kept fixed to the initial value of 2\,$\Omega_{\odot}$ during the disc-locking phase. These initial values are chosen so that the model correctly reproduces the asteroseismic constraints on the core and surface rotation rates obtained for the red giant KIC~7341231 (see the inset in Fig.~\ref{omegas_vs_t_fig}). Such a solution is of course not unique, since an increase (decrease) in the initial angular velocity can be compensated for by an increase (decrease) in the disc lifetime to obtain the same value of total angular momentum on the ZAMS. Both models are computed with shellular rotation only before the post-main-sequence phase and with an additional viscosity $\nu_{\rm add}=1.2 \times 10^{4}$\,cm$^2$\,s$^{-1}$ during the post-main sequence. Figure~\ref{omegas_vs_t_fig} shows that a high degree of differential rotation is already present at the ZAMS for the model that includes shellular rotation during the pre-main sequence. As a result, significant differences between the rotational properties of these models are seen during the first part of the main sequence. The core and surface rotation rates are then found to converge to similar values at the end of the main sequence. This is due to the fact that both models have a similar value of total angular momentum at the ZAMS \cite[see][]{hae13}. More importantly, the rotational properties of both models are nearly identical at the evolutionary stage of KIC~7341231. This shows that the determination of the efficiency of the additional transport mechanism is also almost insensitive to the modelling of the internal angular momentum transport during the pre-main sequence.

\section{Dependance of the additional viscosity on the stellar mass and the evolutionary stage}

\begin{figure}[htb!]
\resizebox{\hsize}{!}{\includegraphics{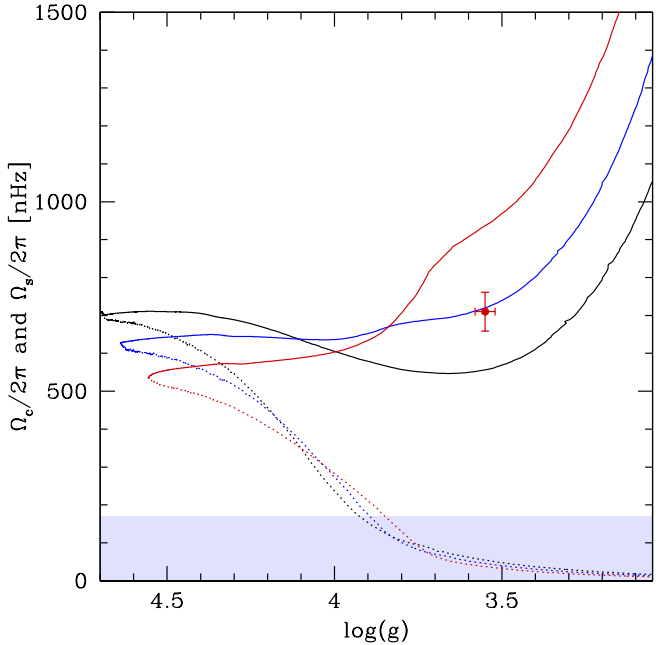}}
 \caption{Core (continuous lines) and surface (dotted lines) rotation rates as a function of gravity for rotating models with different masses. Black, blue and red lines indicate a mass of 0.75, 0.84 and 0.95\,M$_{\odot}$, respectively. All models are computed with an initial velocity on the ZAMS of 2\,km\,s$^{-1}$ and an additional viscosity of $3\times 10^{3}$\,cm$^2$\,s$^{-1}$ ($\nu_{\rm add}$ is included during the whole evolution of the model). The red dot corresponds to the core rotation rate of KIC~7341231 \citep{deh12}.}
  \label{omegacs_vs_logg_mosser_masse}
\end{figure}

The efficiency of the additional viscosity deduced for KIC~7341231 is at least three times lower than the one obtained for KIC~8366239. Being located at the base of the red giant branch, both stars are in a similar evolutionary stage (see Fig.~\ref{dhr_fig}). From core rotation rates deduced for a large sample of red giants by \cite{mos12}, it has been established that, for a given star, the efficiency of the unknown mechanism for angular momentum transport must increase during its evolution on the red giant branch. This has been shown for 1.5\,M$_{\odot}$ models \cite[e.g.][]{can14,egg15} and for a 1.25\,M$_{\odot}$ star \cite[][]{spa16} at solar metallicity. This effect of the evolutionary stage on the efficiency of the additional angular momentum transport process is also found for the low-mass star KIC~7341231. This is illustrated in Fig.~\ref{omegacs_vs_logg_mosser_masse}, which shows the evolution of the core and surface rotation rates as a function of gravity for models of KIC~7341231. With a constant value of $\nu_{\rm add}$, the values of the core rotation rates deduced for stars at the base of the red giant branch can be correctly reproduced, but an increase in the core rotation rate is then predicted when evolution proceeds, in contradiction with the observations of \cite{mos12}. This shows that the efficiency of the transport mechanism increases when the star ascends the red-giant branch. 

\begin{figure}[htb!]
\resizebox{\hsize}{!}{\includegraphics{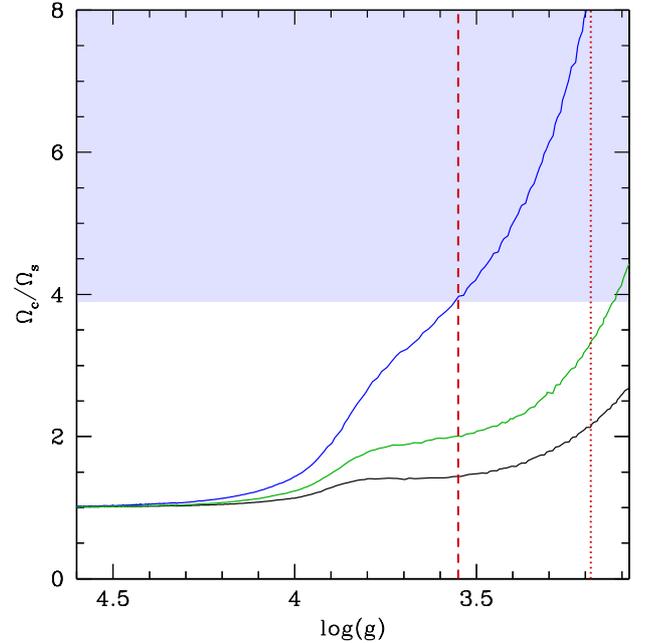}}
 \caption{Ratio of the core-to-surface rotation rate as a function of gravity for models of KIC~7341231. Black, green and blue lines correspond to an additional viscosity of $3 \times 10^{4}$, $2 \times 10^{4}$ and $1 \times 10^{4}$\,cm$^2$\,s$^{-1}$, respectively. The red dashed and dotted vertical lines correspond to the gravity of KIC~7341231 and KIC~8366239, respectively. The blue region indicates the values of the ratio of core-to-surface rotation rate that are compatible with the asteroseismic measurements of KIC~7341231.}
  \label{rap_omegasc_evo_logg}
\end{figure}

While KIC~8366239 and KIC~7341231 are in a similar evolutionary stage, the value of the $\log g$ of the latter is slightly higher than the one of KIC~8366239. Since $\nu_{\rm add}$ seems to increase when $\log g$ decreases, we first investigate whether the difference found between the efficiency of the additional process obtained for both stars cannot be simply explained by this evolutionary effect. We thus follow the evolution of models of KIC~7341231 computed for different values of $\nu_{\rm add}$ until they reach the value of $\log g$ obtained for KIC~8366239. The location of these models in the HR diagram is indicated by an open triangle in Fig.~\ref{dhr_fig}, while the variation in the ratio of core-to-surface rotation rates as a function of $\log g$ is shown in Fig.~\ref{rap_omegasc_evo_logg}. The values of $\log g$ corresponding to KIC~7341231 and KIC~8366239 are indicated by the vertical dashed and dotted lines, respectively. As discussed above, the asteroseismic constraint on the minimum degree of radial differential rotation leads to a maximum value of $\nu_{\rm add}$ of about $1\times 10^{4}$\,cm$^2$\,s$^{-1}$ for KIC~7341231 (blue line in Fig.~\ref{rap_omegasc_evo_logg}). The maximum value of the additional viscosity for a more evolved model of KIC~7341231 that has exactly the same value of $\log g$ than KIC~8366239 (vertical dotted line) is found to be lower than $2\times 10^{4}$\,cm$^2$\,s$^{-1}$ (green line in Fig.~\ref{rap_omegasc_evo_logg}). We thus see that the different values of $\nu_{\rm add}$ derived from asteroseismic measurements of KIC~8366239 ($\nu_{\rm add}=3\times 10^{4}$\,cm$^2$\,s$^{-1}$) and KIC~7341231 are not simply due to a difference in the evolutionary stage of these red giants.

The main differences in the global properties of these stars are the mass (1.5 and 0.84\,M$_{\odot}$ for KIC~8366239 and KIC~7341231, respectively) and the metallicity (solar for KIC~8366239 and $[$Fe/H$] = -1$ for KIC~7341231). To investigate the impact of the stellar mass on the efficiency needed for the additional angular momentum transport process, rotating models of KIC~7341231 are computed with exactly the same initial parameters except for the mass. Figure~\ref{omegacs_vs_logg_mosser_masse} shows the evolution of the core and surface rotation rates as a function of gravity for models of 0.75, 0.84 and 0.95\,M$_{\odot}$ with $\nu_{\rm add} = 3\times 10^{3}$\,cm$^2$\,s$^{-1}$ and an initial velocity on the ZAMS of 2\,km\,s$^{-1}$. An increase in the mass leads to an increase in the core rotation rate (continuous lines in Fig.~\ref{omegacs_vs_logg_mosser_masse}) during the post-main-sequence phase, while surface rotation rates are very similar (dotted lines in Fig.~\ref{omegacs_vs_logg_mosser_masse}). For models with the same $\log g$ as KIC~7341231, an increase in the mass thus results in a more rapidly rotating core together with an increase in the radial differential rotation. A higher efficiency of the internal transport of angular momentum is then required to obtain the same core rotation rate and degree of differential rotation at a given evolutionary stage for more massive models. This shows that in addition to the dependency on the evolutionary stage, the efficiency of the undetermined transport mechanism is also sensitive to the stellar mass. At a given evolutionary stage, the efficiency needed for this transport process is then found to increase with the stellar mass.

\begin{figure}[htb!]
\resizebox{\hsize}{!}{\includegraphics{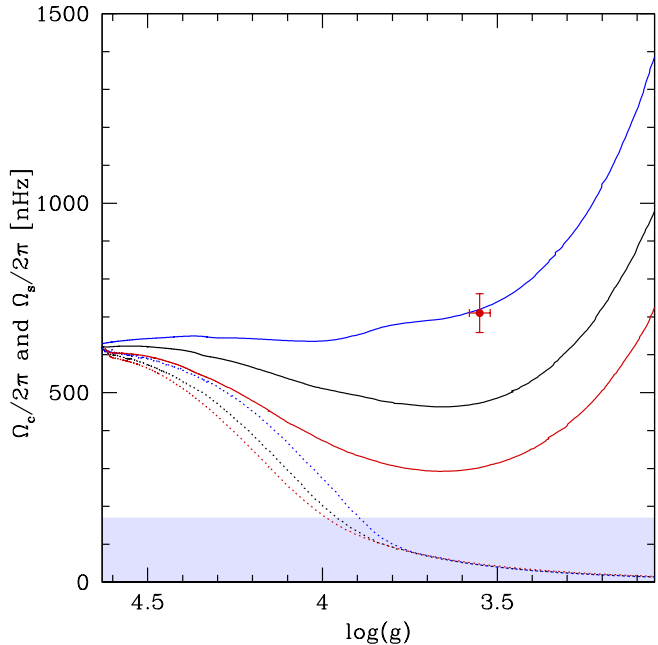}}
 \caption{Same as Fig.~\ref{omegacs_vs_logg_mosser_masse} for models with the same mass of 0.84\,M$_{\odot}$, but different metallicities. Blue, black and red lines correspond to a metallicity [Fe/H]$=-1$, [Fe/H]$=-0.5$, and [Fe/H]$=0$, respectively.}
  \label{omegacs_vs_logg_mosser_feh}
\end{figure}

Another difference between KIC~8366239 and KIC~7341231 is the metallicity, which is solar for KIC~8366239 and equal to $[$Fe/H$] = -1$ for KIC~7341231 \citep{deh12}. One can then wonder whether the different additional viscosities deduced for these stars could also be due to the difference in metallicity. To investigate this point, additional models of KIC~7341231 are computed with metallicities [Fe/H]$=-0.5$ and [Fe/H]$=0$, but with the same mass of 0.84\,M$_{\odot}$, and the same $\nu_{\rm add} = 3\times 10^{3}$\,cm$^2$\,s$^{-1}$ and initial velocity on the ZAMS of 2\,km\,s$^{-1}$. The evolution of the core and surface rotation rates of these models is shown in Fig.~\ref{omegacs_vs_logg_mosser_feh}. An increase in the metallicity leads to a decrease in the core rotation rate at a given value of $\log g$, while surface rotation rates are almost unaffected during the red giant phase. An increase in the metallicity thus results in a lower degree of radial differential rotation; a lower value for the additional viscosity is then needed to obtain a similar value of the core rotation rate and a similar degree of radial differential rotation during the red-giant phase when the metallicity increases. This trend goes in the opposite direction compared to the larger value of $\nu_{\rm add}$ obtained for KIC~8366239 with a solar metallicity. This shows that the difference in metallicity is not responsible for the different values of the additional viscosity obtained for KIC~8366239 and KIC~7341231. This difference is dominated by the sensitivity of $\nu_{\rm add}$ on the stellar mass. We therefore conclude that asteroseismic measurements of KIC~8366239 and KIC~7341231 indicate that the efficiency of the undetermined mechanism for the internal transport of angular momentum increases with the stellar mass.

\section{Conclusion}

We have first characterized the mean efficiency of an undetermined mechanism for the transport of angular momentum in radiative zones that is needed to correctly reproduce the asteroseismic constraints on the core and surface rotation rates of the red giant KIC~7341231 obtained by \cite{deh12}. The mean viscosity $\nu_{\rm add}$ corresponding to this additional transport process is then found to be in the range $1 \times 10^{3}$ -- $1.3 \times 10^{4} $\,cm$^2$\,s$^{-1}$. This relatively large interval of viscosities corresponds to a conservative value for the constraints on $\nu_{\rm add}$, because it includes all the uncertainties on the braking of the stellar surface by magnetized winds, on the efficiency of the internal transport of angular momentun during the pre-main sequence and the main sequence, and on the unknown initial velocity of KIC~7341231. We thus find that the asteroseismic constraints on the internal rotation of KIC~7341231 can provide us with valuable constraints on the efficiency of the internal transport of angular momentum during the post-main sequence independently of the past rotational history of the star. This result is not restricted to the specific case of KIC~7341231 and is valid for any red giant for which a precise core rotation rate and an upper limit on the surface rotation rate can be deduced from asteroseismic measurements of rotational splittings of mixed modes. When one assumes that the additional transport mechanism is already at work during the main sequence together with a solar-calibrated braking of the surface by magnetized winds, the value of $\nu_{\rm add}$ is then restricted to the interval $1$ -- $4 \times 10^{3}$\,cm$^2$\,s$^{-1}$. 

The efficiency of the transport mechanism obtained for KIC~7341231 is found to be lower than the viscosity of $3 \times 10^{4} $\,cm$^2$\,s$^{-1}$ deduced for KIC~8366239. While the efficiency of this additional transport process is known to increase with the evolution of the star during the post-main sequence \cite[e.g.][]{can14,egg15,spa16}, we find that this efficiency also appears to increase with stellar mass. This trend, obtained from a direct comparison between the two red giants KIC~7341231 and KIC~8366239 for which detailed asteroseismic observations are available, is also suggested by the measurements of core rotation rates for a large sample of red giants by \cite{mos12}. We showed that the degree of radial differential rotation and the core rotation rate increase with the stellar mass for models at the same evolutionary stage computed with the same efficiency for the additional transport of angular momentum. As stars with different masses but similar values of core rotation rates can be found at a given evolutionary stage in the sample of \cite{mos12}, this also means that the efficiency of the undetermined angular momentum transport mechanism needs to increase with the stellar mass. 

Interestingly, the increase in the efficiency of the internal transport of angular momentum with the stellar mass obtained for red giants is similar to the trend deduced from observations of surface rotation rates for young low-mass stars in open clusters \cite[see][]{gal15}. Does this mean that the same physical process for the transport of angular momentum in stellar radiative zones is missing during the whole evolution of the star? It is of course difficult to answer such a question without having a clear idea of the physical nature of this mechanism. We showed, however, that the dependence of the efficiency of this transport process on stellar parameters can be characterized and quantified thanks to asteroseismic measurements of red giant stars. This offers a valuable opportunity of revealing its physical nature and of progressing thereby in our global understanding of transport processes in stellar radiative zones.

\begin{acknowledgements}
This work has been supported by the Swiss National Science Foundation grant 200020-160119. NL acknowledges financial support from the CNES postdoctoral fellowship 2016. RAG, TC, and SM acknowledge funding by the European Community Seventh Framework Program (FP7/2007-2013) under the grant agreement 312844 (SPACEINN) and the CNES PLATO grant at CEA-Saclay. SM acknowledges funding by the European Research Council through ERC grant SPIRE 647383. RH acknowledges support from the European Union's Seventh Framework Programme (FP/2007-2013) / ERC Grant Agreement n. 306901 and from the World Premier International Research Center Initiative (WPI Initiative), MEXT, Japan. 
\end{acknowledgements}

%-------------------------------------------------------------------

\bibliographystyle{aa} % style aa.bst
\bibliography{biblio} % your references Yourfile.bib

\begin{thebibliography}{48}
\expandafter\ifx\csname natexlab\endcsname\relax\def\natexlab#1{#1}\fi

\bibitem[{{Beck} {et~al.}(2012){Beck}, {Montalban}, {Kallinger}, {De Ridder},
  {Aerts}, {Garc{\'{\i}}a}, {Hekker}, {Dupret}, {Mosser}, {Eggenberger},
  {Stello}, {Elsworth}, {Frandsen}, {Carrier}, {Hillen}, {Gruberbauer},
  {Christensen-Dalsgaard}, {Miglio}, {Valentini}, {Bedding}, {Kjeldsen},
  {Girouard}, {Hall}, \& {Ibrahim}}]{bec12}
{Beck}, P.~G., {Montalban}, J., {Kallinger}, T., {et~al.} 2012, \nat, 481, 55

\bibitem[{{Belkacem} {et~al.}(2015){Belkacem}, {Marques}, {Goupil}, {Mosser},
  {Sonoi}, {Ouazzani}, {Dupret}, {Mathis}, \& {Grosjean}}]{bel15b}
{Belkacem}, K., {Marques}, J.~P., {Goupil}, M.~J., {et~al.} 2015, \aap, 579,
  A31

\bibitem[{{Benomar} {et~al.}(2015){Benomar}, {Takata}, {Shibahashi},
  {Ceillier}, \& {Garc{\'{\i}}a}}]{ben15}
{Benomar}, O., {Takata}, M., {Shibahashi}, H., {Ceillier}, T., \&
  {Garc{\'{\i}}a}, R.~A. 2015, \mnras, 452, 2654

\bibitem[{{Borucki} {et~al.}(2010){Borucki}, {Koch}, {Basri}, {Batalha},
  {Brown}, {Caldwell}, {Caldwell}, {Christensen-Dalsgaard}, {Cochran},
  {DeVore}, {Dunham}, {Dupree}, {Gautier}, {Geary}, {Gilliland}, {Gould},
  {Howell}, {Jenkins}, {Kondo}, {Latham}, {Marcy}, {Meibom}, {Kjeldsen},
  {Lissauer}, {Monet}, {Morrison}, {Sasselov}, {Tarter}, {Boss}, {Brownlee},
  {Owen}, {Buzasi}, {Charbonneau}, {Doyle}, {Fortney}, {Ford}, {Holman},
  {Seager}, {Steffen}, {Welsh}, {Rowe}, {Anderson}, {Buchhave}, {Ciardi},
  {Walkowicz}, {Sherry}, {Horch}, {Isaacson}, {Everett}, {Fischer}, {Torres},
  {Johnson}, {Endl}, {MacQueen}, {Bryson}, {Dotson}, {Haas}, {Kolodziejczak},
  {Van Cleve}, {Chandrasekaran}, {Twicken}, {Quintana}, {Clarke}, {Allen},
  {Li}, {Wu}, {Tenenbaum}, {Verner}, {Bruhweiler}, {Barnes}, \& {Prsa}}]{bor10}
{Borucki}, W.~J., {Koch}, D., {Basri}, G., {et~al.} 2010, Science, 327, 977

\bibitem[{{Brown} {et~al.}(1989){Brown}, {Christensen-Dalsgaard},
  {Dziembowski}, {Goode}, {Gough}, \& {Morrow}}]{bro89}
{Brown}, T.~M., {Christensen-Dalsgaard}, J., {Dziembowski}, W.~A., {et~al.}
  1989, \apj, 343, 526

\bibitem[{{Cantiello} {et~al.}(2014){Cantiello}, {Mankovich}, {Bildsten},
  {Christensen-Dalsgaard}, \& {Paxton}}]{can14}
{Cantiello}, M., {Mankovich}, C., {Bildsten}, L., {Christensen-Dalsgaard}, J.,
  \& {Paxton}, B. 2014, \apj, 788, 93

\bibitem[{{Ceillier} {et~al.}(2012){Ceillier}, {Eggenberger}, {Garc{\'{\i}}a},
  \& {Mathis}}]{cei12}
{Ceillier}, T., {Eggenberger}, P., {Garc{\'{\i}}a}, R.~A., \& {Mathis}, S.
  2012, Astronomische Nachrichten, 333, 971

\bibitem[{{Ceillier} {et~al.}(2013){Ceillier}, {Eggenberger}, {Garc{\'{\i}}a},
  \& {Mathis}}]{cei13}
{Ceillier}, T., {Eggenberger}, P., {Garc{\'{\i}}a}, R.~A., \& {Mathis}, S.
  2013, \aap, 555, A54

\bibitem[{{Chaboyer} {et~al.}(1995){Chaboyer}, {Demarque}, \&
  {Pinsonneault}}]{cha95}
{Chaboyer}, B., {Demarque}, P., \& {Pinsonneault}, M.~H. 1995, \apj, 441, 865

\bibitem[{{Charbonnel} \& {Talon}(2005)}]{cha05}
{Charbonnel}, C. \& {Talon}, S. 2005, Science, 309, 2189

\bibitem[{{Couvidat} {et~al.}(2003){Couvidat}, {Garc{\'{\i}}a},
  {Turck-Chi{\`e}ze}, {Corbard}, {Henney}, \& {Jim{\'e}nez-Reyes}}]{cou03}
{Couvidat}, S., {Garc{\'{\i}}a}, R.~A., {Turck-Chi{\`e}ze}, S., {et~al.} 2003,
  \apjl, 597, L77

\bibitem[{{Deheuvels} {et~al.}(2015){Deheuvels}, {Ballot}, {Beck}, {Mosser},
  {{\O}stensen}, {Garc{\'{\i}}a}, \& {Goupil}}]{deh15}
{Deheuvels}, S., {Ballot}, J., {Beck}, P.~G., {et~al.} 2015, \aap, 580, A96

\bibitem[{{Deheuvels} {et~al.}(2014){Deheuvels}, {Do{\u g}an}, {Goupil},
  {Appourchaux}, {Benomar}, {Bruntt}, {Campante}, {Casagrande}, {Ceillier},
  {Davies}, {De Cat}, {Fu}, {Garc{\'{\i}}a}, {Lobel}, {Mosser}, {Reese},
  {Regulo}, {Schou}, {Stahn}, {Thygesen}, {Yang}, {Chaplin},
  {Christensen-Dalsgaard}, {Eggenberger}, {Gizon}, {Mathis},
  {Molenda-{\.Z}akowicz}, \& {Pinsonneault}}]{deh14}
{Deheuvels}, S., {Do{\u g}an}, G., {Goupil}, M.~J., {et~al.} 2014, \aap, 564,
  A27

\bibitem[{{Deheuvels} {et~al.}(2012){Deheuvels}, {Garc{\'{\i}}a}, {Chaplin},
  {Basu}, {Antia}, {Appourchaux}, {Benomar}, {Davies}, {Elsworth}, {Gizon},
  {Goupil}, {Reese}, {Regulo}, {Schou}, {Stahn}, {Casagrande},
  {Christensen-Dalsgaard}, {Fischer}, {Hekker}, {Kjeldsen}, {Mathur}, {Mosser},
  {Pinsonneault}, {Valenti}, {Christiansen}, {Kinemuchi}, \&
  {Mullally}}]{deh12}
{Deheuvels}, S., {Garc{\'{\i}}a}, R.~A., {Chaplin}, W.~J., {et~al.} 2012, \apj,
  756, 19

\bibitem[{{Denissenkov} {et~al.}(2010){Denissenkov}, {Pinsonneault},
  {Terndrup}, \& {Newsham}}]{den10_spin}
{Denissenkov}, P.~A., {Pinsonneault}, M., {Terndrup}, D.~M., \& {Newsham}, G.
  2010, \apj, 716, 1269

\bibitem[{{Di Mauro} {et~al.}(2016){Di Mauro}, {Ventura}, {Cardini}, {Stello},
  {Christensen-Dalsgaard}, {Dziembowski}, {Patern{\`o}}, {Beck}, {Bloemen},
  {Davies}, {De Smedt}, {Elsworth}, {Garc{\'{\i}}a}, {Hekker}, {Mosser}, \&
  {Tkachenko}}]{dim16}
{Di Mauro}, M.~P., {Ventura}, R., {Cardini}, D., {et~al.} 2016, \apj, 817, 65

\bibitem[{{Eggenberger}(2015)}]{egg15}
{Eggenberger}, P. 2015, in IAU Symposium, Vol. 307, New Windows on Massive
  Stars, ed. G.~{Meynet}, C.~{Georgy}, J.~{Groh}, \& P.~{Stee}, 165--170

\bibitem[{{Eggenberger} {et~al.}(2005){Eggenberger}, {Maeder}, \&
  {Meynet}}]{egg05_mag}
{Eggenberger}, P., {Maeder}, A., \& {Meynet}, G. 2005, \aap, 440, L9

\bibitem[{{Eggenberger} {et~al.}(2008){Eggenberger}, {Meynet}, {Maeder},
  {Hirschi}, {Charbonnel}, {Talon}, \& {Ekstr{\"o}m}}]{egg08}
{Eggenberger}, P., {Meynet}, G., {Maeder}, A., {et~al.} 2008, \apss, 316, 43

\bibitem[{{Eggenberger} {et~al.}(2010{\natexlab{a}}){Eggenberger}, {Meynet},
  {Maeder}, {Miglio}, {Montalban}, {Carrier}, {Mathis}, {Charbonnel}, \&
  {Talon}}]{egg10_sl}
{Eggenberger}, P., {Meynet}, G., {Maeder}, A., {et~al.} 2010{\natexlab{a}},
  \aap, 519, A116

\bibitem[{{Eggenberger} {et~al.}(2010{\natexlab{b}}){Eggenberger}, {Miglio},
  {Montalban}, {Moreira}, {Noels}, {Meynet}, \& {Maeder}}]{egg10_rg}
{Eggenberger}, P., {Miglio}, A., {Montalban}, J., {et~al.} 2010{\natexlab{b}},
  \aap, 509, A72

\bibitem[{{Eggenberger} {et~al.}(2012){Eggenberger}, {Montalb{\'a}n}, \&
  {Miglio}}]{egg12_rg}
{Eggenberger}, P., {Montalb{\'a}n}, J., \& {Miglio}, A. 2012, \aap, 544, L4

\bibitem[{{Elsworth} {et~al.}(1995){Elsworth}, {Howe}, {Isaak}, {McLeod},
  {Miller}, {New}, {Wheeler}, \& {Gough}}]{els95}
{Elsworth}, Y., {Howe}, R., {Isaak}, G.~R., {et~al.} 1995, \nat, 376, 669

\bibitem[{{Fuller} {et~al.}(2014){Fuller}, {Lecoanet}, {Cantiello}, \&
  {Brown}}]{ful14}
{Fuller}, J., {Lecoanet}, D., {Cantiello}, M., \& {Brown}, B. 2014, \apj, 796,
  17

\bibitem[{{Gallet} \& {Bouvier}(2015)}]{gal15}
{Gallet}, F. \& {Bouvier}, J. 2015, \aap, 577, A98

\bibitem[{{Garc{\'{\i}}a} {et~al.}(2007){Garc{\'{\i}}a}, {Turck-Chi{\`e}ze},
  {Jim{\'e}nez-Reyes}, {Ballot}, {Pall{\'e}}, {Eff-Darwich}, {Mathur}, \&
  {Provost}}]{gar07}
{Garc{\'{\i}}a}, R.~A., {Turck-Chi{\`e}ze}, S., {Jim{\'e}nez-Reyes}, S.~J.,
  {et~al.} 2007, Science, 316, 1591

\bibitem[{{Grevesse} \& {Noels}(1993)}]{gre93}
{Grevesse}, N. \& {Noels}, A. 1993, in Origin and evolution of the elements:
  proceedings of a symposium in honour of H. Reeves, held in Paris, June 22-25,
  1992. Edited by N. Prantzos, E. Vangioni-Flam and M. Casse. Published by
  Cambridge University Press, Cambridge, England, 1993, p.14, ed.
  N.~{Prantzos}, E.~{Vangioni-Flam}, \& M.~{Casse}, 14

\bibitem[{{Haemmerl{\'e}} {et~al.}(2013){Haemmerl{\'e}}, {Eggenberger},
  {Meynet}, {Maeder}, \& {Charbonnel}}]{hae13}
{Haemmerl{\'e}}, L., {Eggenberger}, P., {Meynet}, G., {Maeder}, A., \&
  {Charbonnel}, C. 2013, \aap, 557, A112

\bibitem[{{Kosovichev} {et~al.}(1997){Kosovichev}, {Schou}, {Scherrer},
  {Bogart}, {Bush}, {Hoeksema}, {Aloise}, {Bacon}, {Burnette}, {de Forest},
  {Giles}, {Leibrand}, {Nigam}, {Rubin}, {Scott}, {Williams}, {Basu},
  {Christensen-Dalsgaard}, {D\"appen}, {Rhodes}, {Duvall}, {Howe}, {Thompson},
  {Gough}, {Sekii}, {Toomre}, {Tarbell}, {Title}, {Mathur}, {Morrison}, {Saba},
  {Wolfson}, {Zayer}, \& {Milford}}]{kos97}
{Kosovichev}, A.~G., {Schou}, J., {Scherrer}, P.~H., {et~al.} 1997, \solphys,
  170, 43

\bibitem[{{Krishnamurthi} {et~al.}(1997){Krishnamurthi}, {Pinsonneault},
  {Barnes}, \& {Sofia}}]{kri97}
{Krishnamurthi}, A., {Pinsonneault}, M.~H., {Barnes}, S., \& {Sofia}, S. 1997,
  \apj, 480, 303

\bibitem[{{Lund} {et~al.}(2014){Lund}, {Miesch}, \&
  {Christensen-Dalsgaard}}]{lun14}
{Lund}, M.~N., {Miesch}, M.~S., \& {Christensen-Dalsgaard}, J. 2014, \apj, 790,
  121

\bibitem[{{Maeder}(2009)}]{mae09}
{Maeder}, A. 2009, {Physics, Formation and Evolution of Rotating Stars}
  (Springer Berlin Heidelberg)

\bibitem[{{Maeder} \& {Zahn}(1998)}]{mae98}
{Maeder}, A. \& {Zahn}, J.-P. 1998, \aap, 334, 1000

\bibitem[{{Marques} {et~al.}(2013){Marques}, {Goupil}, {Lebreton}, {Talon},
  {Palacios}, {Belkacem}, {Ouazzani}, {Mosser}, {Moya}, {Morel}, {Pichon},
  {Mathis}, {Zahn}, {Turck-Chi{\`e}ze}, \& {Nghiem}}]{mar13}
{Marques}, J.~P., {Goupil}, M.~J., {Lebreton}, Y., {et~al.} 2013, \aap, 549,
  A74

\bibitem[{{Meynet} {et~al.}(2013){Meynet}, {Ekstrom}, {Maeder}, {Eggenberger},
  {Saio}, {Chomienne}, \& {Haemmerl{\'e}}}]{mey13}
{Meynet}, G., {Ekstrom}, S., {Maeder}, A., {et~al.} 2013, in Lecture Notes in
  Physics, Berlin Springer Verlag, Vol. 865, Lecture Notes in Physics, Berlin
  Springer Verlag, ed. M.~{Goupil}, K.~{Belkacem}, C.~{Neiner},
  F.~{Ligni{\`e}res}, \& J.~J. {Green}, 3--642

\bibitem[{{Mosser} {et~al.}(2012){Mosser}, {Goupil}, {Belkacem}, {Marques},
  {Beck}, {Bloemen}, {De Ridder}, {Barban}, {Deheuvels}, {Elsworth}, {Hekker},
  {Kallinger}, {Ouazzani}, {Pinsonneault}, {Samadi}, {Stello}, {Garc{\'{\i}}a},
  {Klaus}, {Li}, {Mathur}, \& {Morris}}]{mos12}
{Mosser}, B., {Goupil}, M.~J., {Belkacem}, K., {et~al.} 2012, \aap, 548, A10

\bibitem[{{Nielsen} {et~al.}(2015){Nielsen}, {Schunker}, {Gizon}, \&
  {Ball}}]{nie15}
{Nielsen}, M.~B., {Schunker}, H., {Gizon}, L., \& {Ball}, W.~H. 2015, \aap,
  582, A10

\bibitem[{{Palacios} {et~al.}(2006){Palacios}, {Charbonnel}, {Talon}, \&
  {Siess}}]{pal06}
{Palacios}, A., {Charbonnel}, C., {Talon}, S., \& {Siess}, L. 2006, \aap, 453,
  261

\bibitem[{{Pinsonneault} {et~al.}(1989){Pinsonneault}, {Kawaler}, {Sofia}, \&
  {Demarque}}]{pin89}
{Pinsonneault}, M.~H., {Kawaler}, S.~D., {Sofia}, S., \& {Demarque}, P. 1989,
  \apj, 338, 424

\bibitem[{{R\"udiger} \& {Kitchatinov}(1996)}]{rue96}
{R\"udiger}, G. \& {Kitchatinov}, L.~L. 1996, \apj, 466, 1078

\bibitem[{{Spada} {et~al.}(2016){Spada}, {Gellert}, {Arlt}, \&
  {Deheuvels}}]{spa16}
{Spada}, F., {Gellert}, M., {Arlt}, R., \& {Deheuvels}, S. 2016, \aap, 589, A23

\bibitem[{{Spada} {et~al.}(2010){Spada}, {Lanzafame}, \& {Lanza}}]{spa10}
{Spada}, F., {Lanzafame}, A.~C., \& {Lanza}, A.~F. 2010, \mnras, 404, 641

\bibitem[{{Spruit}(1999)}]{spr99}
{Spruit}, H.~C. 1999, \aap, 349, 189

\bibitem[{{Spruit}(2002)}]{spr02}
{Spruit}, H.~C. 2002, \aap, 381, 923

\bibitem[{{Tayar} \& {Pinsonneault}(2013)}]{tay13}
{Tayar}, J. \& {Pinsonneault}, M.~H. 2013, \apjl, 775, L1

\bibitem[{{Turck-Chi{\`e}ze} {et~al.}(2010){Turck-Chi{\`e}ze}, {Palacios},
  {Marques}, \& {Nghiem}}]{tur10}
{Turck-Chi{\`e}ze}, S., {Palacios}, A., {Marques}, J.~P., \& {Nghiem}, P.~A.~P.
  2010, \apj, 715, 1539

\bibitem[{{van Saders} {et~al.}(2016){van Saders}, {Ceillier}, {Metcalfe},
  {Silva Aguirre}, {Pinsonneault}, {Garc{\'{\i}}a}, {Mathur}, \&
  {Davies}}]{van16}
{van Saders}, J.~L., {Ceillier}, T., {Metcalfe}, T.~S., {et~al.} 2016, \nat,
  529, 181

\bibitem[{{Zahn}(1992)}]{zah92}
{Zahn}, J.-P. 1992, \aap, 265, 115

\end{thebibliography}

\end{document}